\documentclass[sigconf]{acmart}

\usepackage{xcolor}         
\usepackage{algorithm}
\usepackage{multirow}
\usepackage{float}
\usepackage{subcaption}
\usepackage{multirow}
\usepackage{float} 
\usepackage{wrapfig}
\usepackage{color}
\usepackage{graphics}
\usepackage{hyperref}

\AtBeginDocument{%
  \providecommand\BibTeX{{%
    \normalfont B\kern-0.5em{\scshape i\kern-0.25em b}\kern-0.8em\TeX}}}

\copyrightyear{2023}
\acmYear{2023}
\setcopyright{acmlicensed}\acmConference[WWW '23]{Proceedings of the ACM Web Conference 2023}{May 1--5, 2023}{Austin, TX, USA}
\acmBooktitle{Proceedings of the ACM Web Conference 2023 (WWW '23), May 1--5, 2023, Austin, TX, USA}
\acmPrice{15.00}
\acmDOI{10.1145/3543507.3583383}
\acmISBN{978-1-4503-9416-1/23/04}

\begin{document}

\title{
Few-shot News Recommendation via Cross-lingual Transfer}

\author{Taicheng Guo}
\affiliation{%
  \institution{University of Notre Dame}
  \city{Notre Dame}
  \country{USA}
  }
\authornote{The work was mainly completed when the first author was a Master student at King Abdullah University of Science and
Technology, Saudi Arabia.}
\email{tguo2@nd.edu}

\author{Lu Yu}
\affiliation{%
  \institution{Ant Group}
  \city{Hangzhou}
  \country{China}}
\email{bruceyu.yl@alibaba-inc.com}

\author{Basem Shihada}
\affiliation{%
  \institution{King Abdullah University of Science and
Technology}
  \city{Thuwal}
  \country{Saudi Arabia}
  }
\email{basem.shihada@kaust.edu.sa}

\author{Xiangliang Zhang}
\affiliation{%
  \institution{University of Notre Dame}
  \city{Notre Dame}
  \country{USA}}
\email{xzhang33@nd.edu}
\authornotemark[0]
\authornote{Corresponding author. Xiangliang Zhang is secondly affiliated with King Abdullah University of Science and Technology, Saudi Arabia }


\begin{abstract}

The cold-start problem has been commonly recognized in recommendation systems and studied by following a general idea to leverage the abundant interaction records of warm users to infer the preference of cold users. 
However, the performance of these solutions is limited by the amount of records available from warm users to use. 
Thus, building a recommendation system based on few interaction records from  a few users   
still remains a challenging problem for unpopular or early-stage recommendation platforms. 
This paper focuses on solving the \emph{few-shot recommendation} problem for news recommendation  based on two observations. First, news at different platforms (even in different languages) may share similar topics. Second,  
the user preference over these topics is transferable across different platforms.
Therefore, we propose to solve the \emph{few-shot news recommendation} problem by 
transferring the user-news preference  from a  many-shot source domain to a few-shot target domain.  
To bridge two domains that are even in different languages and without any overlapping users and news, we propose a novel unsupervised cross-lingual transfer model as the news encoder that aligns semantically similar news in two domains. 
A user encoder is constructed on top of the aligned news encoding and transfers the user preference from the source to target domain. 
Experimental results on two real-world news recommendation datasets show  the superior performance of our proposed method on addressing few-shot news recommendation, comparing to the   baselines.
The source code can be found at \href{https://github.com/taichengguo/Few-shot-NewsRec}{\underline{https://github.com/taichengguo/Few-shot-NewsRec.}}
\end{abstract}

\ccsdesc[500]{Information systems~Recommender systems}

\keywords{News recommendation; Cross domain recommendation; Transfer learning}


\maketitle

\section{Introduction}
News recommendation  aims to  help users find  news  that they are interested in over the massive options~\cite{10.1145/1754239.1754257, Phelan11termsof, wu-etal-2019-neural-news}. 
Such personalized recommender systems boost users' reading experience and business revenue of news platforms~\cite{10.1145/3530257, wu-etal-2021-newsbert-distilling}. 
It has thus garnered increasing attention and has been tackled by a number of  methods~\cite{wu-etal-2020-mind, Wu2019NPA, ijcai2019-0536, wu-etal-2019-neural-news}. 
{The general idea of these methods is to characterize the preference of a user based on the news that the user has previously read and then infer the interest of the user on other news. The quality of recommendation results thus highly depend on the amount of historical user-news  interactions to use. Like all other recommendation problems, news recommendation also suffers from the cold-start problem when the   targeted user only has   interactions with few news before, as illustrated in Fig. \ref{fig:1}(a). 

\begin{figure}[t]
  \Description{Illustration of  problem formulations and solutions. (a) Cold-start problem: predict the preference of  cold-start users while having many warm users in the system. (b) Few-shot problem: predict the preference of cold-start users while no warm users are available. (c)/(d) Solving few-shot problem by cross-domain transfer with/without overlapping users between two domains.}
  \begin{subfigure}{0.2\textwidth}
    \includegraphics[width=0.9\linewidth]{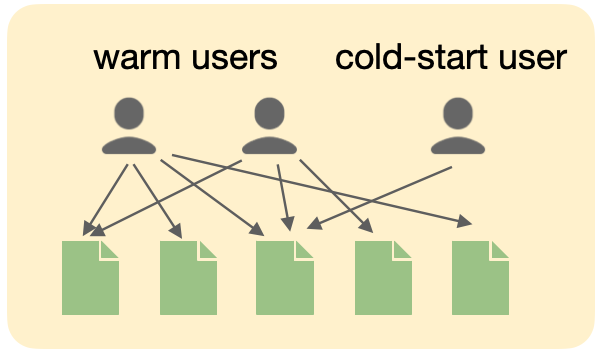}\vspace{-0.14cm}
       \caption{Cold-start problem}
     \label{fig:1a}
  \end{subfigure}%
  \hspace*{\fill}   
  \begin{subfigure}{0.2\textwidth}
    \includegraphics[width=0.9\linewidth]{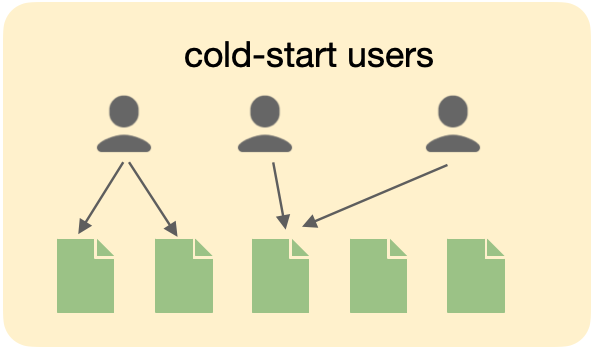}\vspace{-0.14cm}
      \caption{Few-shot problem}
     \label{fig:1b}
  \end{subfigure}%
  
  \vspace{+0.2cm}
  \begin{subfigure}{0.4\textwidth}
  \centering
    \includegraphics[width=0.8\linewidth]{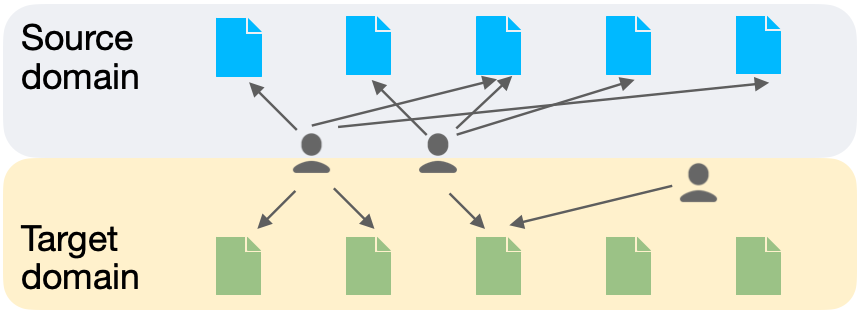}\vspace{-0.14cm}
      \caption{Solving few-shot problem  by cross-domain transfer with overlapping users}
     \label{fig:1c}
    \end{subfigure}
    
   \vspace{+0.2cm}
     \begin{subfigure}{0.4\textwidth}
     \centering
    \includegraphics[width=0.8\linewidth]{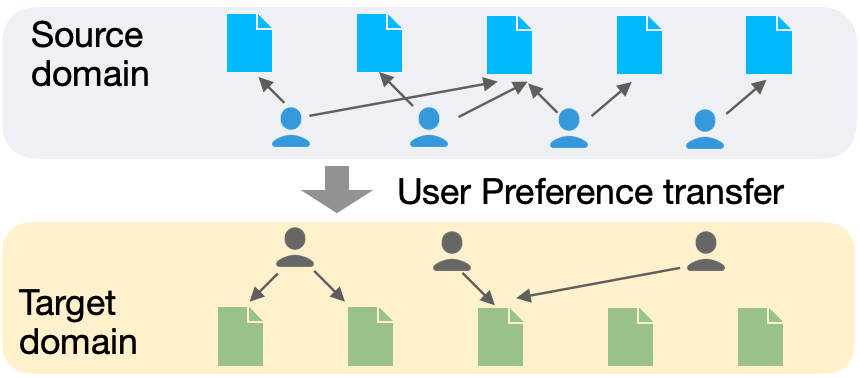}\vspace{-0.14cm}
       \caption{Solving few-shot problem  by cross-domain transfer without overlapping users  (ours)}
     \label{fig:1d}
  \end{subfigure}
\vspace{-0.2cm}
\caption{Illustration of  problem formulations and solutions. (a) Cold-start problem: predict the preference of  cold-start users while having many warm users in the system. (b) Few-shot problem: predict the preference of cold-start users while no warm users are available. (c)/(d) Solving few-shot problem by cross-domain transfer with/without overlapping users between two domains. 
} \label{fig:1} \vspace{-0.3cm}
\end{figure}

A number of works have tried  to infer the preference of cold-start users based on the behavior of warm
users who have browsed a large amount of news \cite{KARIMI20181203, Personalizednewsimplicitsocialexperts, PENETRATE, zeroShotRec}}.
Though  promising results have been obtained by making use of warm users to help cold users in recommendation, these solutions become incapable  when no warm users are available to use. This is the so-called \emph{few-shot user recommendation} scenario, where all users have few user-item interactions, as shown in Fig. \ref{fig:1}(b). 
{This scenario can often be found at unpopular or early-stage recommendation platforms. } 
Addressing the few-shot recommendation at these platforms is  actually a  chicken-and-egg problem.  The performance of the recommender system at these platforms relies on the observable user-item interactions and in turn, the user-item interactions rely on the performance of the recommender system as users may not find their interested items and leave the platform~\cite{ding2021zeroshot}.
To address this issue in news recommendation,  an approach in \cite{Cold-Start_News14} is proposed to use  the browse information of these cold-start users themselves in other domains.
This is a feasible idea to transfer the knowledge from a rich source domain to the few-shot target domain.
However, these methods require the existence of overlapping users between two domains \cite{ijcai2017-343, hu-yang-2021-trnews,Cold-Start_News14} (as shown in Fig. \ref{fig:1}(c)), and are thus limited to only transfer   knowledge via the overlapping users.




We focus on a more challenging but widely existing problem that   users in a target domain are all cold-start users, and they are not available in other domains for playing the role of knowledge transfer. As shown in   Fig. \ref{fig:1}(d), these users have only a few interaction records in one single domain. In this case, the previous meta-learning based   \cite{FewShotLBSN, MFNP, MELU} and transfer learning based solutions  \cite{ijcai2017-343, hu-yang-2021-trnews,Cold-Start_News14} become incapable, due to the limited amount of user-news interactions. Even a strong domain-specific news encoder cannot always help in this case, because the inference of users' interest is  limited more by the lack of user-news interactions, rather than the understanding of news content. Take zero-shot cold users as special examples, a news encoder doesn't help on guessing the users' interest. Recommendation decision is then likely made by 
blind guess, unless external information from another domain can help on profiling the common patterns of users' interest. 

\begin{figure}[t]
\Description{Overlapping news category distribution between   Adressa (in Norwegian) and MIND (in English) dataset.}
\begin{center}
   \includegraphics[width=1\linewidth]{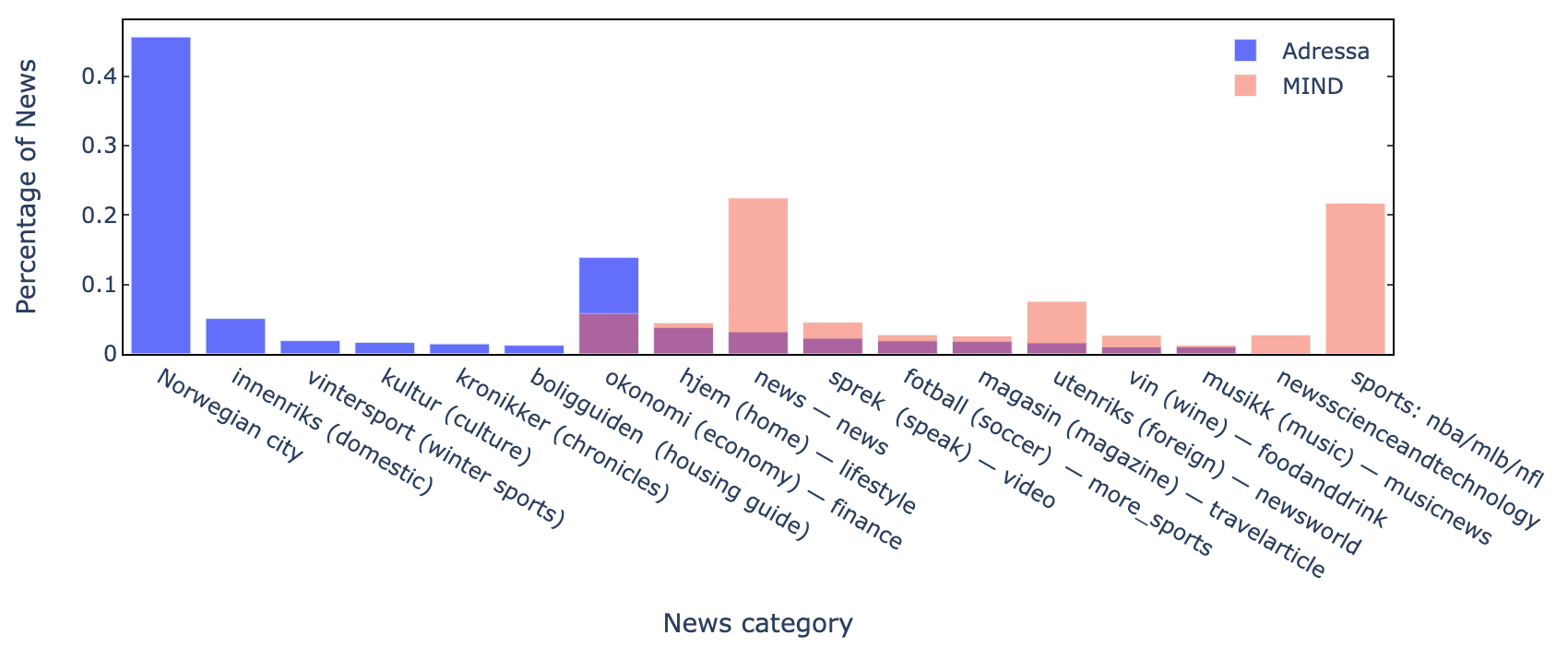}
\end{center}
\label{fig:long} \vspace{-0.2cm}
\caption{Overlapping news category distribution between   Adressa (in Norwegian) and MIND (in English)
dataset.} \vspace{-0.2cm}
\label{cate}
\end{figure}

We are therefore motivated to build a few-shot cross-domain news recommendation system, which works without the requirement of  overlapping users to bridge the two domains. Picturing our application on two widely used public news recommendation datasets: MIND (English news  published around 2020) and Adressa (Norwegian news  published around 2017), we have two observations that support our investigation of the cross-domain (even cross-lingual) transfer.
First, news in different domains may share similar categories and have somewhat similar 
topics. Figure \ref{cate} shows that the news categories in MIND  and Adressa dataset have some overlapping topics, such as general \emph{news},  \emph{financial news}, \emph{home/lifestyle} related news and so on. Second, although users in two domains are different, some general reading preference over news topics is transferable from the domain of one language to anther.

Based on these observations, we propose to build a \emph{few-shot cross-lingual news recommendation system}. Since there are no overlapping users or news to bridge the two domains, transferring the user preference patterns   from one domain to another has to address  the \emph{domain shift issue}, which   can be attributed to  {the difference of news category distribution between two domains. As shown in Figure \ref{cate}, the news category distributions of two datasets are different, although there are overlapping topics. The news about \emph{Norwegian city} is the majority in Adressa, while \emph{sport news} and general \emph{news} are the two major topics in MIND. Since the preference of users should be inferred over the news distribution in their own domain,  it is therefore essential to minimize the gap of user characterization   between the source and target domain. When the target domain is cold for all users, it is a new challenging problem for addressing the domain shift issue by only a few user-news interactions.  The previous study  of language shift in  NLP tasks like name entity recognition  \cite{zhang-etal-2021-cross, johnson-etal-2019-cross}}  {
developed methods that align different language text with similar topic distribution and are  not suitable for solving the gap of user characterization between domains in recommendation.  Multilingual pretrained language models can be employed to represent  news of different languages in the same space. However, the representation only takes into account the news content, and cannot catch how they are liked by users. In news recommendation models, news should be represented to reflect not only its content but also its attractiveness to users.  } 

We design a cross-domain recommendation system consisting of  news and user encoder shared between the source and target domain and propose three strategies for alleviating the domain shift, \emph{Cross-domain Extension}, \emph{Random Masking}, and \emph{News-Alignment}. The 
\emph{Cross-domain Extension} strategy is to construct augmented news which are semantically similar across two domains. 
The \emph{Random Masking} strategy works by randomly taking the original news  or one of  the augmented news  during training. 
In this way,  {the training of the shared news encoder is exposed to both the source and target domain. The inference of user preference later based on the news encoder thus covers the topics in target domain as well.
}
The \emph{News-Alignment} strategy is designed to drive similar news from two domains to be close  in the same representation space. It presents  all augmented news in pairs, and sends them to train the  shared news encoder, such that the paired news have similar representations. 
In this common news representation space, we then  get the user representations by the shared user encoder  based on users’ browsing history. The recommendation problem in target domain is thus solved with the transferred knowledge from the source domain.

We summarize the contributions of this work as follow:
(1) To the best of our knowledge, we are the first one to use the cross lingual knowledge to solve the few-shot news recommendation problem in a challenging setting of no common users or news existing between the source and target  domain. 
(2) We propose to use a shared recommender model across two domains to transfer the user-news preference patterns. For combating with the domain shift  issue,  we design three strategies to align the source and target domain in the same representation space.  
(3) We demonstrate the effectiveness of our proposed method with thorough experiments on two real-world  datasets. The results  
show that our  method has a consistent improvement above the   baselines.


\section{Related Work}
\textbf{Neural News Recommendation:} 
On the recent benchmark  news recommendation dataset MIND~\cite{wu-etal-2020-mind}, various
deep learning-based news recommendation architectures have been developped, such as NPA~\cite{Wu2019NPA}, NAML~\cite{ijcai2019-0536}, and NRMS~\cite{wu-etal-2019-neural-news}.  
Pre-trained language models~\cite{wu-etal-2021-newsbert-distilling} and even multilingual pre-trained models 
~\cite{10.1145/3404835.3463069} have been used as advanced news encoder for improving the recommendation performance. 
Besides, cold-start issues have been also investigated in news recommendation by inferring the preference of cold users based on their activities in other domains \cite{hu-yang-2021-trnews,Cold-Start_News14} or based on activities of other warm users in the same domain \cite{KARIMI20181203, Personalizednewsimplicitsocialexperts, PENETRATE, zeroShotRec}. 

\textbf{Cold-start and Few-shot User Recommendation:} 
In the general recommendation field, including news recommendation, cold-start  problems for user refer to the situation where little is known about the preferences of the new user~\cite{KARIMI20181203}. Most solutions~\cite{Cold-Start_News14, KARIMI20181203, Personalizednewsimplicitsocialexperts, PENETRATE, zeroShotRec, hu-yang-2021-trnews} aim at leveraging the behaviors of warm users to  infer the preference of cold-start users. However, these methods become incapable in the scenario where no warm users are available  and all users have few interactions with items. For this  \emph{few-shot user recommendation} problem,  
a popular solution is to apply   meta-learning   on making  recommender models   adaptable to users who have few  interactions~\cite{FewShotLBSN, MFNP, MELU}. However, for many unpopular or early-stage recommendation platforms, these meta-learning based solutions become incapable due to the lack of training data for constructing meta-learning tasks. In ~\cite{ding2021zeroshot},   zero-shot recommenders are proposed to leverage the knowledge of a source dataset to improve the recommendation performance of a target domain without using any target domain data during training. 
Our proposed solution is flexible for either zero-shot or few-shot setting, depending on if any training data are available or not in the target domain. 

\textbf{Cross Domain Recommendation:} 
Previous cross domain recommendation models~\cite{ijcai2017-343, hu-yang-2021-trnews} are mainly designed for the cold-start problems, i.e., the user or item is new. For bridging the two different domains, they assume the existence of overlapping users or news between two domains. 
Our study is in a realist setting that has no requirement on the existence of overlapping users or news. 
The study of domain adaptation recommendation~\cite{Yu2020a} also doesn't require overlapping users or items. However, the adaptation methods require extensive training data in both domains and thus are not applicable for the few-shot scenario where the training samples in the target domain are scarce.  

\textbf{Cross-lingual Transfer and Multilingual Pretrained Language Models:} 
Cross-lingual techniques have promoted various NLP tasks. 
One popular idea is  to translate the source language text to the target language. Based on the translated source text, the translation-then-align method \cite{li2021unsupervised, zhang-etal-2021-cross, klinger-cimiano-2015-instance}, Bilingual method \cite{li2021unsupervised, zhang-etal-2021-cross} 
and many other task-specific cross-lingual methods~\cite{johnson-etal-2019-cross, zhang-etal-2021-cross} have been designed to reduce the language shift between two domains.
These aforementioned works are designed to transfer the text-level knowledge between languages which are in the same topics for text-level tasks, e.g., the user reviews sentiment analysis about similar topics in different languages in SemEval-2016 task~\cite{pontiki-etal-2016-semeval}. For the few-shot news recommendation task, each user is represented by a sequence of reading history news and the user preference  is inferred based on the sequence of news. Because of the news topics deviation, the target domain reading history has a  different topic distribution from the source domain reading history. The inference of user preference in a cross-lingual model  thus should address the domain shift issue. 
Multilingual pretrained language models can be an effective technique to find news with similar content  in different languages. {They can be employed as a news encoder for representing  news from different domains in the same space. However, the news will be encoded solely based on the content, without considering the relevance of news in users' reading history. Two pieces of news with different content and topics can have similar representation if users who read one also prefer to read the other.  Hence, the news encoder in a recommendation model should represent news by taking into account both its content and its history of being read by users. }

\section{The Proposed Few-shot News Recommendation Model}
\subsection{Notation and Problem Definition}  
We let the source domain be noted as $s$ and the target domain   be noted as $t$. The $U^s$, $U^t$ and $D^s$, $D^t$ represent the user and news set of source domain and target domain, respectively. The user-news interactions in source and  target domain available for training are noted as $\tau^s$ and $\tau^t$, respectively. 
Each user $u \in U$ has a reading history news sequence $[d_1, d_2, ..., d_{len(u)}]$, where $len$ measures the length of a sequence or the size of a set. Each news $d \in D$ 
is represented by a token sequence $[w_1, w_2, ..., w_{len(d)}]$. 
For a  given user  $u \in U$ and a candidate news set $C=\{d_i\},i=1...|C|$, 
the recommendation task is to predict the preference score of this user on each candidate news: $r_i$, $i=1...|C|$.
In the few-shot recommendation problem, we have only the ground-truth label $y_i$ for a few news that the user $u$ likes ($y_i$=1) or not ($y_i$=0).
We build a cross-lingual news recommendation system, which  predicts the preference of user $u \in U^t$ in the target domain  by leveraging the abundant user-news interactions $\tau^s$ in the source domain and the few-shot or zero-shot interactions $\tau^t$ in the target domain.   
\subsection{Overall Framework}

\begin{figure*}[tp]
\Description{Framework of our proposed method.}
\begin{center}
\includegraphics[scale=0.3]{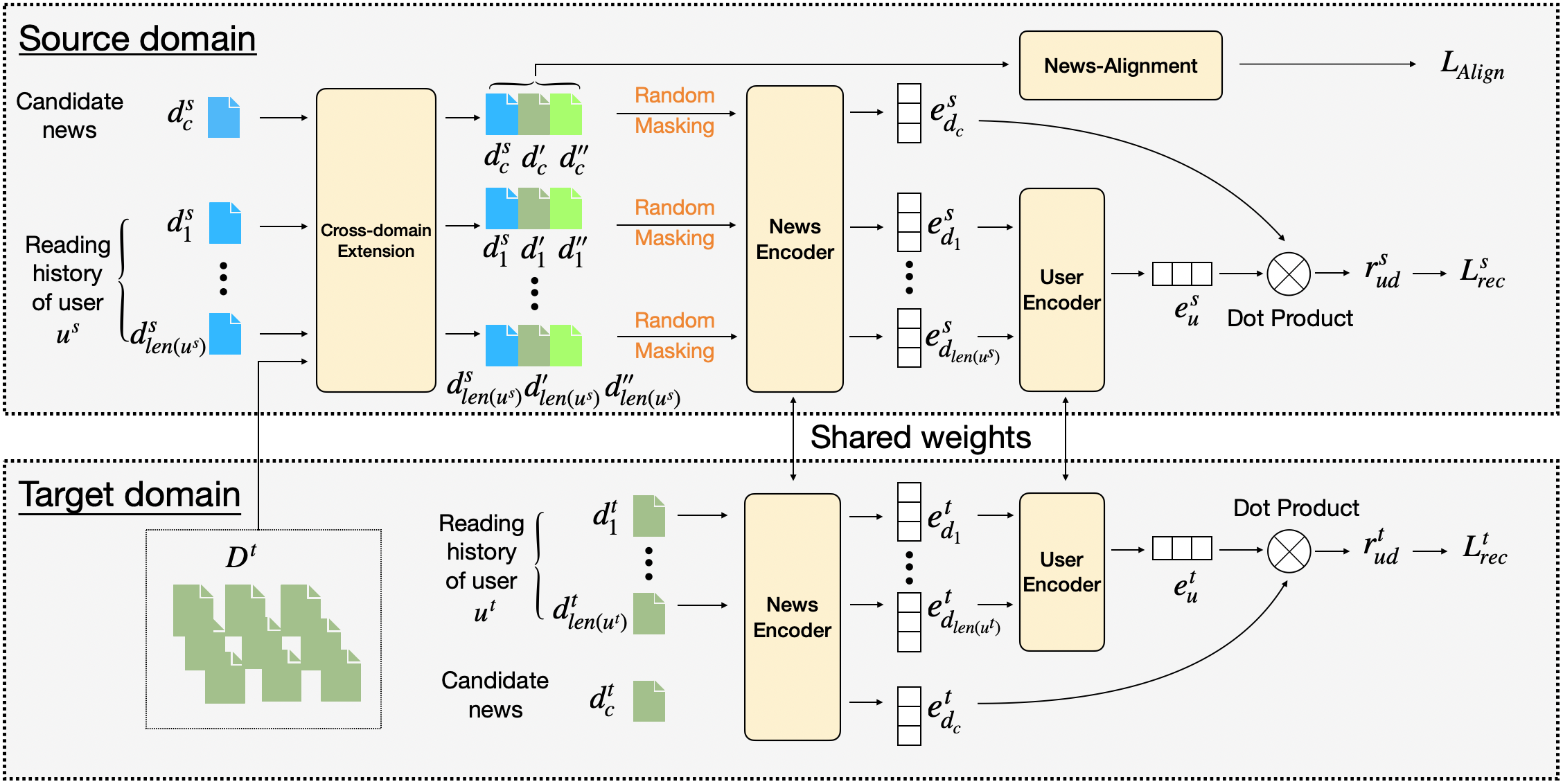}
\end{center}\vspace{-0.2cm}
\caption{Framework of our proposed method.}
\label{framework}
\end{figure*}

The framework of our proposed method is shown in Figure \ref{framework}. 
The shared news encoder and user encoder are jointly trained by the source domain training set  $\tau^s$ and the target domain training set  $\tau^t$. Note that $\tau^t$ is much smaller than $\tau^s$, and there is no overlapping between $\tau^s$ and $\tau^t$.
After training, the news and user encoder are used in the target domain for predicting $r^t_{ud}$, the preference of user $u$ on news $d$ in target domain.  
There is a module of Cross-domain Extension (CDE) in training. For a news in the source domain $d \in D^s$, it goes through the CDE module and gets $d'$ (the $d$'s  most similar news from the target domain)
and  $d''$ (a translation of $d$ in the target language). Such extension  is designed to alleviate the  language shift and content shift for training the shared news encoder after Random Masking (RM). 
The RM operation 
randomly masks two news among ($d$,  $d'$,  $d''$) and sends the remaining one to news encoder for participating the training. In the iterative training process, this RM operation blends the augmented news in the original news and makes the shared 
news encoder exposed to both the source and target domain.
the source and target domain be better aligned.  
The news encoding vectors  are then sent to user encoder to produce user encoding vectors. The preference prediction $r^s_{ud}$ and $r^t_{ud}$ are calculated by running dot product on  the encoding vectors of candidate news and user. One additional News-Alignment module is applied to the source domain news and the augmented news for  driving the source and target domain into the same representation space, so the domain shift can be further alleviated. 

\subsection{Base Recommendation Network}
We first describe the base  neural recommendation model for  obtaining users representations, news representations and  constructing the recommendation loss to train the neural network model. We choose the state-of-the-art model NRMS~\cite{wu-etal-2019-neural-news}  as our base network. The source domain and target domain share the same base network to encode news and users, and then conduct recommendations.

\textbf{News Encoder $\phi$}: Our news encoder is shared by the source and target domain. A news $d$ from either source or target domain  is first tokenized by bpe~\cite{heinzerling2018bpemb} and then sent  to a multilingual shared dictionary to obtain the representations of each token $[e_{w_1}, e_{w_2},…,e_{w_{len(d)}}]$. Then the token representations are given to a multi-head self-attention and attention network  to produce the news representation $e_d$,
\begin{equation}
e_d = \phi(d) = Attention(MultiHeadAttn([e_{w_1}, e_{w_2},…,e_{w_{len(d)}}])).
\end{equation}

\textbf{User Encoder $\varphi$}: The user representation is built from the reading history $[d_1, d_2, ..., d_{len(u)}]$ of this user. Thus, the news representation $\phi(d_i)$ in the history is sent to an attention network to produce, 
\begin{equation}
    e_u = \varphi(u) = Attention([\phi(d_1), \phi(d_2),…,\phi(d_{len(u)})]).
\end{equation}

\textbf{Training and Inference}: Given the representation of a user and a candidate news, the preference value can be computed by dot product,    $r_{ud} = e_u^Te_d$.
In the training stage, we have a positive news $d$ liked by   a user $u$, and sample four random negative news that the user may not like.
We thus have   $[d^+, d_1^-, d_2^-, d_3^-, d_4^-]$ and its corresponding prediction score $[r_{ud}^+, r_{u1}^-, r_{u2}^-, r_{u3}^-, r_{u4}^-]$. 
The recommendation loss can be defined by soft-max on  the predicted scores. 
The overall loss function defined on training data is 
\begin{equation}
  \mathcal{L}_{rec} = -\sum_{(u, C) \in \tau}{\sum_{i \in C_{positive}}{\log \frac{\exp \left(r_{ui}^{+}\right)}{\exp \left(r_{ui}^{+}\right)+\sum_{j=1}^{4} \exp \left(r_{uj}^{-}\right)}}},
\end{equation}
where $positive(C)$ is the positive news among the candidate news.

\subsection{Cross-domain Extension}
The key of designing a cross-lingual news recommendation system is to address the \emph{domain shift issue}, which can be attributed to the difference of news  distribution between two domains.
{
Although multilingual pre-trained language models can represent text of different languages in the same embedding space and alleviate the language shift to certain extend, they are mostly trained only with token-level  objectives to align tokens in different languages~\cite{lin-etal-2021-common, 10.5555/3454287.3454921}. If employing them as news encoder, news will be encoded solely based on the content, without considering the relevance of news in terms of users’ preference. In the task of news recommendation, news representation should be learned based on not only its content but also its attractiveness to users. }

To address this domain shift issue, we introduce 
the news of the target domain into the source domain during  training  and  simulate the interactions of source users in the target domain and also the interaction of target users in the source domain. This so-called  \emph{Cross-domain Extension} (CDE) module is our first strategy to alleviate the domain shift via augmenting news across two domains. Specifically, there are two ways of augmentation.
First, we  use a multilingual pre-trained language model to find the most similar news $d' \in D^t$ in the target domain for each news $d \in D^s$ in the source domain:
$$d^{\prime} = FindMostSimilar(d, D^t).$$ {Concretely, we first use Sentence Transformers ~\cite{reimers-2019-sentence-bert} and Multilingual DistilBertModel ~\cite{reimers-2020-multilingual-sentence-bert} to obtain the embeddings for each news in the source domain and target domain. Based on the embeddings of news, we use Hnswlib~\cite{malkov2018efficient} which is a fast approximate nearest neighbor search tool to  find the  most similar news  $d' \in D^t$ for each $d \in D^s$. To exclude  unrelated news in nearest neighbor search, we set a similarity threshold ranging from 0 to 1 to control the selection process. For each $d \in D^s$, if its similarity score to the nearest neighbor is still lower than the threshold, it fails to find a similar news in the target domain and will have no $d^{\prime}$ in augmentation.
We  explore the influence of the similarity threshold in   Section 4.}

{
To weaken the  language shift between two domains, we translate each  news in the source domain to
the language in target domain.} Machine translation has been regarded as an effective technique to rewrite input text to a semantic similar translated text whose grammar and syntax are more similar to the target domain.
Formally, for each news in the source domain $d \in D^s$, we use the Google translate\footnote{https://translate.google.com.} to translate it into the target language $$d^{\prime\prime} = Translate(d).$$ News $d$ and $d^{\prime\prime}$
 will be the two forms of expression about the same news content in different languages. 
 
{The \emph{Cross-domain Extension} is operated before training and can be regarded as a pre-processing method and run offline.}
The augmented news set of $d$, denoted as $A_d = \{d, d', d''\}$, 
will be used in the next \emph{Random Masking} module for simulating the interactions of users with news across different domains. 

\subsection{Document-level Random Masking, News and User Encoding}



{After introducing the news of the target domain into training data, we design \emph{Document-level Random Masking}, the second strategy for reducing the domain shift. The goal is to blend the augmented news in the original news for training the shared news encoder by making it exposed to both the source and target domain. 
Concretely, 
}
the random masking operation is applied to $A_d = \{d, d', d''\}$ by randomly masking two of them and sending the remaining one to the news encoder (for news $d$ that doesn't have a similar target news, randomly mask one of $\{d, d''\}$ and send the remaining one to the news encoder).
The idea is analogous to  code-switch used in cross-lingual NLP tasks~\cite{ijcai2020-0533, zhang-etal-2021-cross}, which randomly replaces words in a sentence with the corresponding translation words in a different language to align words in different languages by mixing their context information. We propose a similar document-level operation in the recommendation task for simulating the interactions of
users with news across different domains. 

Imagine that one training sample in the source domain includes a user $u \in U^s$ with reading history $[d_1^s, d_2^s, ..., d^s_{len(u)}]$, a candidate news set $C$ 
and the label for each candidate news which indicates the user likes or dislike the news. For each news in the training sample (including both the history and candidate set), we have its augmented set $A_{d_i} = \{d_i, d'_i, d''_i\}$. In the iterative training process, the random masking operation will dynamically create enriched reading history and candidate news set for user $u$, which is a mixture of news in source and target domain. 
Let $\Gamma$ be the random masking function. We then get the news representation for $d_i$: $$e_{d_i} = \phi(\Gamma(A_{d_i})),$$  and the user representation for $u$:  $$e_u = \varphi(u) = Attention([\phi(\Gamma(A_{d_1})), \phi(\Gamma(A_{d_2})),…,\phi(\Gamma(A_{d_{len(u)}}))]).$$

In this way, 
users in the source domain are virtually interacting with news in both source and target domain. This random masking operation is also beneficial for users in the target domain. If a  user $u^t \in U^t$ interacted with news $d' \in D^t$ in history, the virtual interaction between $u^s \in U^s$ with this $d' \in D^t$ will pull $u^t$ and $u^s$ closer. Meanwhile, the news $d \in D^s$ and its translated $d''$ will be pulled close to $u^t$ as well. The target user $u^t$ is thus virtually interacting with news in the two domains.  The shared news encoder $\phi$ and user encoder $\varphi$ trained by these augmented user-news interactions  reduce the 
deviation of news distribution between the source and target domain, and therefore have promoted generalization capacity when recommending news to cold users in the target domain. 
Note that there is no distortion of user preference in the random masking, because there is no deletion of the original user-news interactions, 
and $d'_i$ and $d''_i$ are derived from $d_i$ to have similar content.  
\subsection{News-Alignment}
To ensure that the  news with similar content in one augmented news set $A_d = \{d, d', d''\}$ are embedded as representation vectors with high similarities,  we design a news-alignment module, which explicitly pulls $\{d, d', d''\}$ close in the representation space. 
For each news $d \in D^s$, we firstly get its augmented news set $A_d =\{d, d', d''\}$, and then we use the news encoder to get representation of the source domain news $e_d = \phi(d)$, the translated news $e_d' = \phi(d')$, and the target domain news $e_d'' = \phi(d'')$. 
The mean squared error loss (MSE) below is minimized to align $\{d, d', d''\}$ closer in the representation space,
\begin{equation}
L_{Align}=\sum_{d \in D^s} MSE(e_d, e_d') + MSE(e_d, e_d'').
\end{equation}
In this way, the news commonly interacted by users are further pulled closer, no matter the  interactions are original or augmented, are in source or target domain. Even when the translated news $d''$ has an issue of translation quality, the alignment module helps to adjust the news encoder $\phi$ so the embedding of $d''$ does not deviate from that of $d$.


The overall loss function for the cross-domain new recommendation model consists of the recommendation loss in the source and target domain, as well as  the alignment loss: 
\begin{equation}
L = \alpha L_{Align} + \beta L_{rec}^s + L_{rec}^t
\end{equation}
where $\alpha$ and $\beta$ are the trade off parameters. 
Our model is flexibly applicable to few-shot and zero-shot setting.  In a zero-shot setting, there will be no target domain recommendation loss.

\section{Experiment Evaluation}
\subsection{Experiment settings}
\paragraph{\textbf{Datasets}}
There are five public news recommendation datasets including Plista (German) ~\cite{Plista}, Adressa (Norwegian)\footnote{https://reclab.idi.ntnu.no/dataset with CC BY-NC-SA 4.0 license.}~\cite{10.1145/3106426.3109436}, Globo (Portuguese)~\cite{globo}, Yahoo!(English)\footnote{https://webscope.sandbox.yahoo.com/catalog.php?datatype=l} and MIND (English)\footnote{https://msnews.github.io with Microsoft Research License Terms.}~\cite{wu-etal-2020-mind}. The Globo and Yahoo! datasets only have  word embeddings or word IDs for the news content. Since the  original text of news is not available, we cannot get the multilingual embedding of news in these datasets, and thus cannot use them in the evaluation. The Plista dataset is not
yet public available until the submission of this work. Hence, there are only two public news recommendation datasets available for our evaluation.
We use MIND (English) and Adressa (Norwegian) to conduct our experiments. MIND is a news recommendation dataset released by Microsoft in 2020. It includes massive user-news interactions in MSN news website. Adressa is a news recommendation dataset released by NTNU in 2017. It includes Norwegian news and user-news interactions in Norwegian on the platform Adresseavisen which is a news website in Trondheim, Norway.

\paragraph{\textbf{Few-shot (zero-shot) transfer setting.} }
We construct two few-shot transfer settings: Adressa -> MIND and MIND -> Adressa.  
In both transfer settings, the training samples in the target domain are prepared in the few-shot scenario by randomly selecting 200 users, each of which only has two news interactions (2-shot), or four news interactions (4-shot). The zero-shot scenario has no training samples in the target domain. 
To imitate the real-world few-shot scenario, the news in  selected trainig samples were  published before the date of test set. 
More details about the training and test dataset can be found in Table \ref{tab1}. 
In MIND -> Adressa, we randomly selected 10,000 users and their news interactions from the training set of MINDlarge as the source domain training set. We randomly selected 40\% of all users and their interactions in Adressa on Jan. 7, 2017 as the validation set and the other 60\% as the test set in the target domain.
In Adressa -> MIND, we randomly selected 10,000 Adressa users who click news on January 6, 2017 to construct the training set in source domain, and include also the reading history of these users collected from January 1, 2017 to January 5, 2017. For the target domain dataset, we randomly selected 40\% of users and their interactions in dev set of MINDlarge as the validation set and the other 60\% as the test set. 

\paragraph{\textbf{Implementation details and evaluation metrics}}
We implement our model by Pytorch and conduct all the experiments on a Linux server with GPUs (Nvidia RTX 3090). We use the Adam optimizer for training. The training epochs are set to 20, and the embedding dimension is set to 300, and the batch size is set to 80. We  use AUC as the main evaluation metric. The performance measured by MRR, NDCG@5, NDCG@10 is also reported in the Appendix. In the MIND -> Adressa experiment, the learning rate is set to 3e-4. In the Adressa -> MIND experiment, the learning rate is set to 1e-4. For both settings, we performed grid search in \{0, 0.2, 0.4, 0.6, 0.8, 1.0\} for the hyper-parameter $\alpha$ and $\beta$. The best $\alpha$ is set to 1 and $\beta$ is set to 0.2, indicating the necessity of $\alpha$ to align news, and $\beta$ to transfer knowledge and prevent news embedding overfitted to source domain. Each experiment is repeated for 5 times and the mean and the standard deviation are reported.
\begin{table*}[]
\Description{Dataset statistics in the two few-shot (and zero-shot) transfer settings (Note that  the test set in target domain is larger than the training set in the few-shot news recommendation scenario)}
\caption{Dataset statistics in the two few-shot (and zero-shot) transfer settings (Note that  the test set in target domain is larger than the training set in the few-shot news recommendation scenario).} \vspace{-0.3cm}
\begin{tabular}{ccc|ccc|ccc}
\hline
\multicolumn{3}{c|}{\multirow{2}{*}{Setting}}                                                                                                                                                                   & \multicolumn{3}{c|}{MIND -> Adressa}                   & \multicolumn{3}{c}{Adressa -> MIND}                     \\ \cline{4-9} 
\multicolumn{3}{c|}{}                                                                                                                                                                                           & \multicolumn{1}{l|}{0-shot} & \multicolumn{1}{c|}{2-shot} & 4-shot & \multicolumn{1}{l|}{0-shot} & \multicolumn{1}{c|}{2-shot} & 4-shot \\ \hline
\multicolumn{1}{c|}{\multirow{5}{*}{\begin{tabular}[c]{@{}c@{}}Training\\ Set\end{tabular}}} & \multicolumn{1}{c|}{\multirow{2}{*}{\begin{tabular}[c]{@{}c@{}}Target\\ Domain\end{tabular}}}  & \#users         & \multicolumn{1}{c|}{0}      & \multicolumn{1}{c|}{200}    & 200    & \multicolumn{1}{c|}{0}      & \multicolumn{1}{c|}{200}    & 200    \\ \cline{3-9} 
\multicolumn{1}{c|}{}                                                                        & \multicolumn{1}{c|}{}                                                                          & \#interactions  & \multicolumn{1}{c|}{0}      & \multicolumn{1}{c|}{400}    & 800    & \multicolumn{1}{c|}{0}      & \multicolumn{1}{c|}{400}    & 800    \\ \cline{2-9} 
\multicolumn{1}{c|}{}                                                                        & \multicolumn{1}{c|}{\multirow{3}{*}{\begin{tabular}[c]{@{}c@{}}Source \\ Domain\end{tabular}}} & \#users         & \multicolumn{3}{c|}{10000}                                         & \multicolumn{3}{c}{10000}                                          \\ \cline{3-9} 
\multicolumn{1}{c|}{}                                                                        & \multicolumn{1}{c|}{}                                                                          & \#news          & \multicolumn{3}{c|}{31099}                                         & \multicolumn{3}{c}{2097}                                           \\ \cline{3-9} 
\multicolumn{1}{c|}{}                                                                        & \multicolumn{1}{c|}{}                                                                          & \#interactions  & \multicolumn{3}{c|}{1083619}                                       & \multicolumn{3}{c}{22700}                                          \\ \hline
\multicolumn{2}{c|}{\multirow{2}{*}{Test set}}                                                                                                                                                & \#users         & \multicolumn{1}{l|}{2194}   & \multicolumn{1}{c|}{816}    & 837    & \multicolumn{1}{l|}{1000}   & \multicolumn{1}{c|}{864}    & 843    \\ \cline{3-9} 
\multicolumn{2}{c|}{}                                                                                                                                                                         & \#interactions & \multicolumn{1}{l|}{4674}   & \multicolumn{1}{c|}{1734}   & 1777   & \multicolumn{1}{l|}{1627}   & \multicolumn{1}{c|}{1380}   & 1278   \\ \hline
\end{tabular}
\label{tab1}
\end{table*}

\subsection{Baselines}
There is no method specifically designed for cross-lingual few-shot news recommendation. We   adopt  common cross-domain  and domain-adaption recommendation methods as baselines:\\
\textbf{NRMS (Only Target data)}~\cite{wu-etal-2019-neural-news}: This is our base model. In zero-shot setting, we evaluate it on the target domain test set directly. In few-shot setting, we train the model on the target domain training set and then evaluate it on the target domain test set.\\ 
\textbf{ZERO-SHOT (Source + Target data)}~\cite{zhang-etal-2021-cross}: 
We adapt the framework of cross-lingual models designed for sentiment analysis in \cite{zhang-etal-2021-cross} as  baselines for our recommendation tasks.
In zero-shot setting, a recommendation model is trained on the source data and tested on the target domain test data. 
In the few-shot setting, the target domain training set is included in training as well.   Testing is always on   the target domain test set.  
\\
\textbf{Translate-then-align (Translation + Target data) / (Source data + Translation)}~\cite{zhang-etal-2021-cross}: Following the Translate-then-align method in \cite{zhang-etal-2021-cross}, news samples in the source language are translated into the target language by machine translation. Then the translated dataset are used for training in the zero-shot setting. In the few-shot setting,   the target domain training set is also included for training. We also add another baseline where the model is trained on the source language, and the news samples in the target language are translated into the source language for inference. \\
\textbf{Bilingual (Source + Translation + Target data)}~\cite{zhang-etal-2021-cross}: As a variant of \textbf{Translation + Target data},
this baseline uses both the translated dataset and the original dataset of source domain for training   in the zero-shot setting. In the few-shot setting, the target domain training set is additionally included for training. \\ 
\textbf{TDAR}~\cite{Yu2020a}: This is a text-enhanced domain adaptation recommendation model.  
It requires training data in both source and target domain. We thus only have  it for comparison in the few-shot setting.



\begin{table*}[]
\Description{Zero-shot and few-shot recommendation performance comparison in terms of AUC. 
The best result is in bold font, and the best  baseline is underlined. The unsolvable zero-shot case is indicated by $-$.}
\centering
\caption{Zero-shot and few-shot recommendation performance comparison in terms of AUC. 
The best result is in bold font, and the best  baseline is underlined. The unsolvable zero-shot case is indicated by $-$.} \vspace{-0.3cm}
\resizebox{\textwidth}{!}{
\begin{tabular}{ll|lll|lll}
\hline
\multicolumn{2}{l|}{Cross-domain Setting}                                           & \multicolumn{3}{c|}{MIND -> Adressa}                                                                                        & \multicolumn{3}{c}{Adressa -> MIND}                                                                                          \\ \hline
\multicolumn{2}{l|}{training data available in  target domain}        & 0-shot                                  & 2-shot                                  & 4-shot                                 & 0-shot                                  & 2-shot                                  & 4-shot                                  \\ \hline
\multicolumn{2}{l|}{NRMS (Only Target)~\cite{wu-etal-2019-neural-news}}                          & 0.482$\pm$0.012          & \underline{0.564$\pm$0.006}          & 0.542$\pm$0.004         & 0.506$\pm$0.012          & 0.507$\pm$0.014          & 0.490$\pm$0.016           \\
\multicolumn{2}{l|}{ZERO-SHOT (Source + Target)~\cite{zhang-etal-2021-cross}}                 & \underline{0.512$\pm$0.014}          & 0.549$\pm$0.018          & 0.529$\pm$0.010          & 0.522$\pm$0.014          & 0.528$\pm$0.021          & 0.520$\pm$0.014          \\
\multicolumn{2}{l|}{Trans.-Align (Trans. + Target)~\cite{zhang-etal-2021-cross}}      & 0.504$\pm$0.016           & 0.551$\pm$0.009          & 0.538$\pm$0.015         & \underline{0.523$\pm$0.004}          & \underline{0.533$\pm$0.012}          & 0.509$\pm$0.015          \\        
\multicolumn{2}{l|}{Trans.-Align (Source + Trans.)~\cite{zhang-etal-2021-cross}}    & 0.495$\pm$0.010          & 0.529$\pm$0.010         & 0.534$\pm$0.007         & 0.522$\pm$0.009          & \underline{0.533$\pm$0.003}          & \underline{0.526$\pm$0.012}         \\ 
\multicolumn{2}{l|}{Bilingual (Source+Trans.+Target)~\cite{zhang-etal-2021-cross}}    & 0.505$\pm$0.014          & 0.539$\pm$0.017          & \underline{0.546$\pm$0.014}         & 0.518$\pm$0.014          & 0.528$\pm$0.005          & 0.513$\pm$0.012          \\ 
\multicolumn{2}{l|}{TDAR~\cite{Yu2020a}}                                        &$-$                      & 0.509$\pm$0.010           & 0.507$\pm$0.013         &  $-$                   & 0.498$\pm$0.012          & 0.515$\pm$0.014          \\ \hline
\multicolumn{1}{c|}{\multirow{2}{*}{Ours}} & Random Masking            & 0.521$\pm$0.011          & 0.570$\pm$0.016           & 0.574$\pm$0.009         & \textbf{0.542$\pm$0.005} & 0.535$\pm$0.004          & \textbf{0.532$\pm$0.011} \\
\multicolumn{1}{c|}{}                      & Random Masking+News-Align & \textbf{0.536$\pm$0.014} & \textbf{0.586$\pm$0.017} & \textbf{0.588$\pm$0.020} & 0.540$\pm$0.005           & \textbf{0.537$\pm$0.008} & 0.531$\pm$0.006          \\ \hline
\end{tabular}
}
\label{tab2:auc}
\end{table*}

\subsection{Overall recommendation performance in zero-shot and few-shot setting }
The overall recommendation performance measured by AUC is reported in Table \ref{tab2:auc}. The results of other metrics are presented in Table \ref{tab2} in Appendix A.1. 
From the  results  shown in Table \ref{tab2:auc} and \ref{tab2},  we have the following observations:\\
\textbf{1)} Our  method  outperforms all baselines in all settings. In the results of zero-shot setting  in Table \ref{tab2:auc}, our method achieves   4.69\% AUC improvement over the best baseline in MIND -> Adressa and 3.63\% in Adressa -> MIND. In the 2-shot and 4-shot setting, the AUC improvements over the best baseline are 3.9\% and 7.69\% in MIND -> Adressa, 0.75\% and 2.31\% in Adressa -> MIND, respectively. \\
\textbf{2)} The inferior performance of ZERO-SHOT (Source + Target) indicates that
fine-tuning model on the source language and then inferring on the target language is not an effective method for knowledge transfer because the transfer is restricted to the implicit domain alignment from the pre-training process as we discussed in the Related Work section. 
The Translate-then-align and Bilingual methods are worse than our methods, due to  the language shift and content shift issue we mentioned in the Introduction section. \\
\textbf{3)} TDAR has low performance in the low-resource cross lingual news recommendation scenario. This is mainly because   TDAR works better when  topics overlap more between two domains. The evaluation data in the two domains were published in different periods and thus have  content shift. Therefore, it is difficult for   TDAR to align the content distribution between two domains.

\begin{figure}
\Description{User embeddings visualization of ZERO-SHOT (Source + Target) method and our method in the MIND -> Adressa 4-shot setting.}
\begin{center}
   \includegraphics[width=1\linewidth]{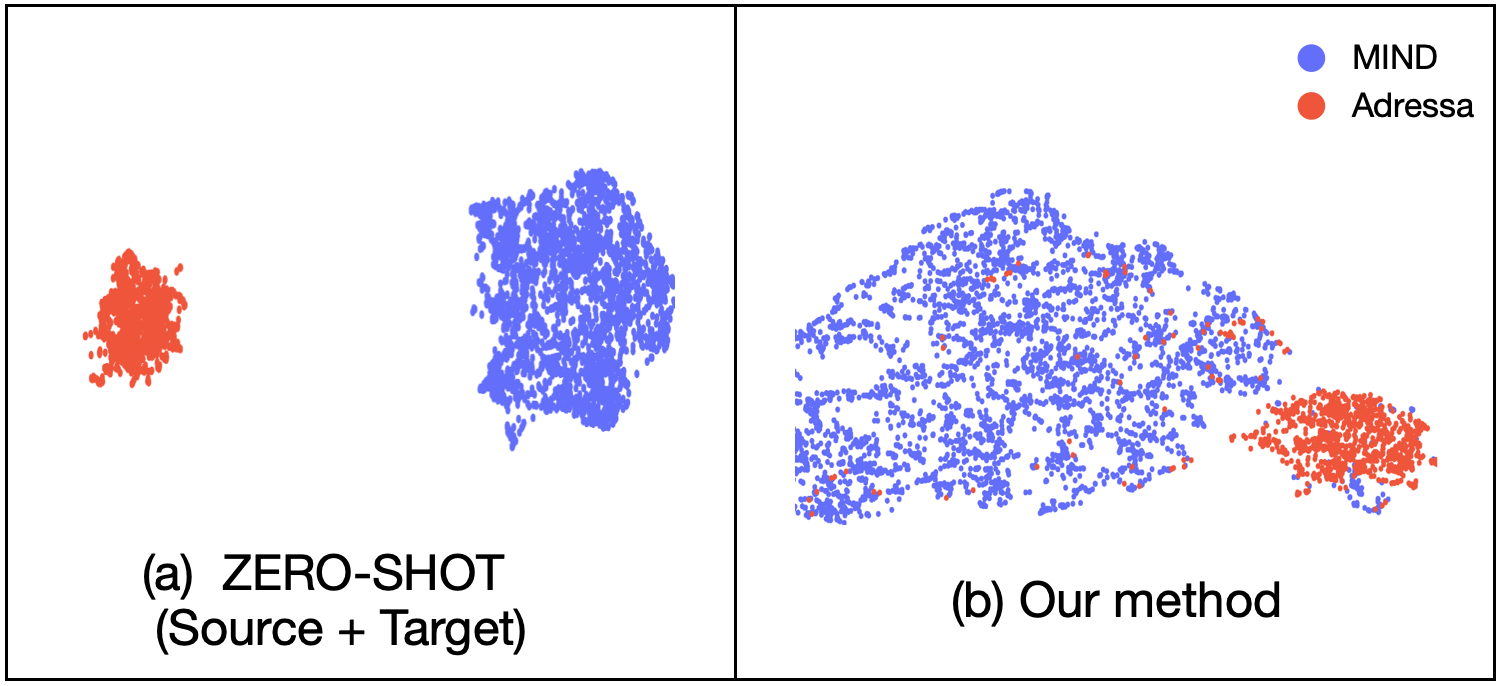}
\end{center} \vspace{-0.2cm}
\caption{User embeddings visualization of ZERO-SHOT (Source + Target) method and our method in the MIND -> Adressa 4-shot setting. }
\label{embedding}\vspace{-0.4cm}
\end{figure}

To demonstrate the capability of our model on alleviating the domain shift when characterizing user preference, we use UMAP~\cite{mcinnes2018umap-software} to reduce users embedding of two domains to 2-dimensional space and then visualize. As shown in Figure \ref{embedding}, comparing to the baseline  ZERO-SHOT (Source + Target), our method pulls close user embedding vectors of two domains. There also exist users in one domain but embedded close to users in another domain, i.e., the red (blue) points in the region of blue (red) points.  Our method is thus verified to be able to transfer user preference patterns, even without requiring  bridging users/news commonly exist in two different domains.   

It is worth to discuss interesting findings from the experiment results in two different transfer settings (MIND -> Adressa vs Adressa -> MIND): the improvements in the MIND -> Adressa setting are all better than the improvements in the Adressa -> MIND setting. This can be explained by the property of source/target domain.  MIND contains a large amount of English news covering a wide range of topics, 
while Adressa only contains a small number of Norweigian news and  limited topics in Norway. Thus, in the MIND -> Adressa setting, the source domain  contains more user preference knowledge of the target domain, comparing to the case in the  Adressa -> MIND transfer setting. So the improvements are  more significant.
In real-world scenarios, the MIND -> Adressa transfer setting is also more realistic than the  Adressa -> MIND  setting, because most news websites are  English. The abundant user-news interaction records from English websites can be used to solve the few-shot problem in other low-resource news recommendation platforms.  
\subsection{Ablation studies }

\begin{table}[]
\Description{The impact of Cross-domain Extension, Random Masking and News-Alignment, evaluated in the MIND -> Adressa 4-shot setting.}
\centering
\caption{The impact of Cross-domain Extension, Random Masking and News-Alignment, evaluated in the MIND -> Adressa 4-shot setting.}\vspace{-0.3cm}
\scalebox{0.95}{
\begin{tabular}{ll|l}
\hline
\multicolumn{2}{l|}{Method}                                                                                                                     & AUC            \\ \hline
\multicolumn{2}{l|}{Bilingual (Source+Trans.+Target)~\cite{zhang-etal-2021-cross}(Best Baseline)}                                                                                                                       & 0.546$\pm$0.014         \\ \hline
\multicolumn{1}{l|}{\multirow{3}{*}{\begin{tabular}[c]{@{}l@{}}without \\ target \\ news   $d^{\prime}$\\\end{tabular}}} & Random Masking                  & 0.558$\pm$0.011 \\
\multicolumn{1}{l|}{}                                                                                         & News-Alignment                  & 0.559$\pm$0.009 \\
\multicolumn{1}{l|}{}                                                                                         & Random Masking + News-Alignment     & 0.569$\pm$0.007 \\ \hline
\multicolumn{1}{l|}{\multirow{3}{*}{\begin{tabular}[c]{@{}l@{}} with \\ target  \\ news $d^{\prime}$\\\end{tabular}}}   & Random Masking                  & 0.574$\pm$0.009         \\
\multicolumn{1}{l|}{}                                                                                         & News-Alignment                  & 0.559$\pm$0.009 \\ 
\multicolumn{1}{l|}{}                                                                                         & Random Masking + News-Alignment & \textbf{0.588$\pm$0.020}         \\ \hline
\end{tabular}
}
\label{abstudy}\vspace{-0.3cm}
\end{table}

To further investigate the effect of three strategies designed for reducing domain shift, we conduct ablation studies in the MIND -> Adressa 4-shot setting. The   results are shown in Table \ref{abstudy}. 

\textbf{The use of target domain news $d' \in D^t$ in Cross-domain Extension}.
From Table \ref{abstudy}, we can observe that all models with target domain news $d'$ in Cross-domain Extension have better performance than those without it, indicating the selected target domain news in Cross-domain Extension can be regarded as an effective bridge to help  transfer  useful knowledge from the source domain to the target domain. 

\textbf{Random Masking and News-Alignment}.
In both "with target news" and "without target news" setting, the performance of Random Masking and News-Alignment are better than the best baseline. The performance of Random Masking + News-Alignment are the best. This demonstrates that each module is beneficial to the model and they can compensate with each other and help model achieve better performance.


\subsection{Experiments with different amounts of users from the target domain}
We also evaluate the impact on transfer performance when varying the amounts of few-shot users from the target domain in training. 
In the MIND -> Adressa setting, we randomly sample 500, 1500, 2500, 3500, 4500 users who have less than 10 interactions in the target domain as the few-shot training samples and sample 8000 users as the test set. We select two strong baselines for comparison and the experiment results are shown in Figure \ref{user_increase} (more results are presented in Appendix A.2). We find that our method consistently achieves better performance, especially when the number of users in training from the target domain is 500, 1500, and 2500. With the increase of  training users, the improvements over the baselines decrease. This is an expected trend because domain transfer is less useful when more training data are available in target domain.

\begin{figure}[t]
\Description{Performance when varying the amount of users involved in training from the target domain.}
\vspace{0.1cm}
\begin{center}
   \includegraphics[width=0.95\linewidth]{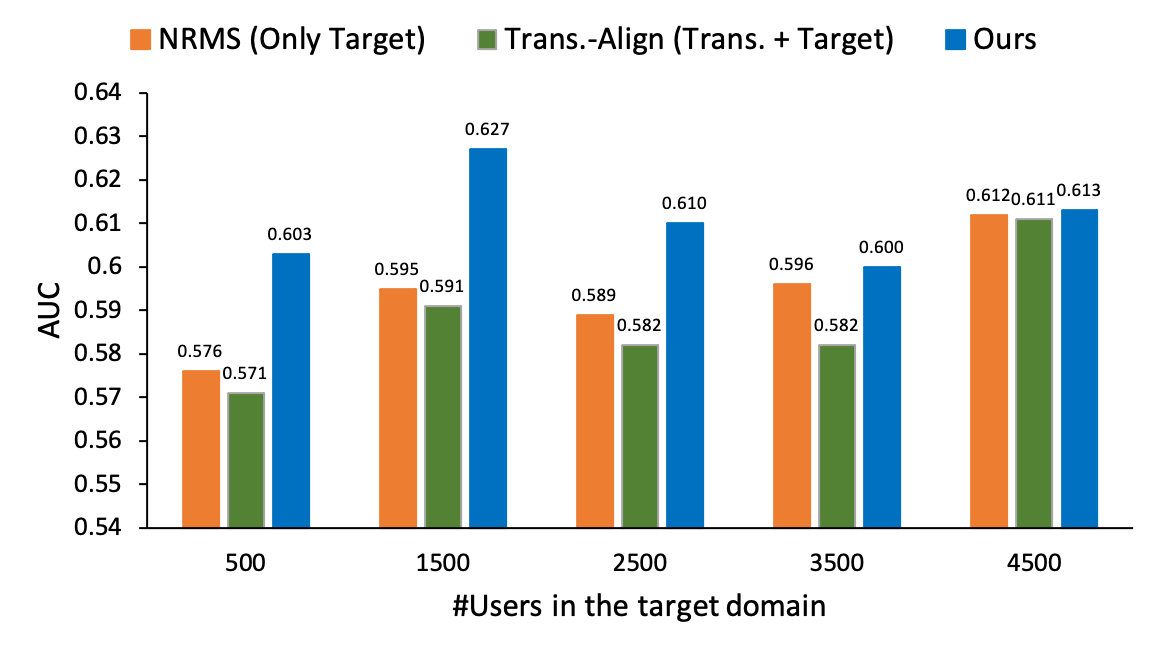}
\end{center}
\vspace{-0.6cm}
\caption{Performance when varying the amount of users involved in training from the target domain.}
\label{user_increase}\vspace{-0.3cm}
\end{figure}

\subsection{Experiments with the multilingual pretrained language model}

\begin{figure}[t]
\Description{Performance when using the Multilingual Pretrained Language Model as news encoder.}
\vspace{0.1cm}
\begin{center}
   \includegraphics[width=0.95\linewidth]{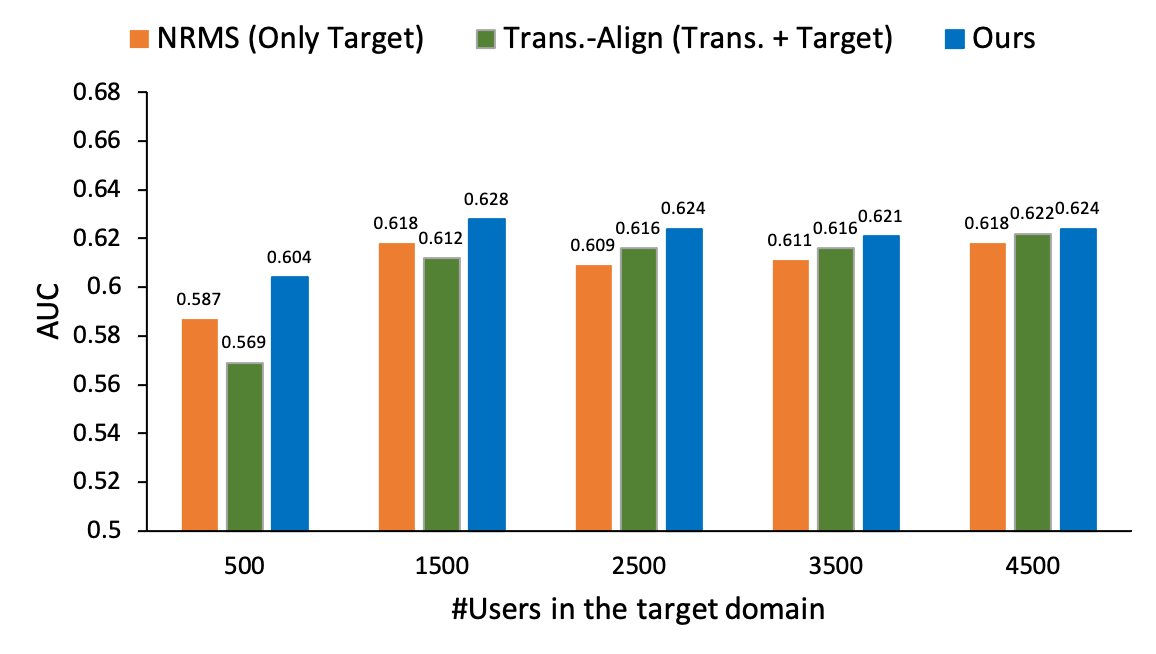}
\end{center}
\vspace{-0.6cm}
\caption{Performance when using the Multilingual Pretrained Language Model as news encoder.}
\label{plm}\vspace{-0.4cm}
\end{figure}

The above reported experiments are from our method using
 NRMS~\cite{wu-etal-2019-neural-news}  as the base recommendation model. In fact, our method is flexible for replacing the news encoder by any other language models. 
To further explore whether our strategies of reducing domain shift work well with other news encoders, we employ the multilingual pretrained language model - MultilingualBert~\cite{devlin-etal-2019-bert} as news encoder.
The training of news encoder will be based on the same strategies but just fine-tuning the language model parameters. We use MultilingualBert also for the NRMS and Trans.-Align(Trans.+Target) baseline. The performance comparison in terms of AUC is shown in Figure \ref{plm}. More results are presented in Appendix A.3.
We can see that for all methods in most settings, using the pretrained language model as the news encoder can relatively achieve better performance than the previous news encoder in NRMS. 
Comparing Figure \ref{user_increase} and \ref{plm}, we can see that two baseline models benefit more from the multilingual news encoder than our method. However, our method can still outperform the baselines, at all different amounts of target domain users in  training.

\section{Conclusion}
Few-shot news recommendation is a challenging and practical task for many unpopular or early-stage platforms. In this paper, we firstly   solve this problem by cross-lingual transfer. 
To address the key challenge of domain shift in building a cross-lingual news recommendation model, we employ a shared news and user encoder between two domains to help transfer user-news preference patterns. We  design three components which  align two domains  to reduce the domain shift. Based on existing news datasets, we construct two few-shot transfer settings and conduct extensive experiments. Experiment results show our model can outperform the baselines. 
A broader impact of our work is to help build recommender systems for countries or regions with low-resource languages.
However, since our model is trained offline and may not be able to work well for breaking news that's never been seen before in neither the source and target domain.   

\begin{acks}
The research reported in this paper was supported by funding from King Abdullah University of Science and Technology (KAUST).
\end{acks}

\bibliographystyle{ACM-Reference-Format}
\bibliography{sample-base}

\appendix
\section{Influence of the similarity threshold}

\begin{figure}[!htbp]
\Description{Influence of the similarity threshold  for selecting the most similar target domain news.}
\vspace{0.1cm}
\begin{center}
   \includegraphics[width=1\linewidth]{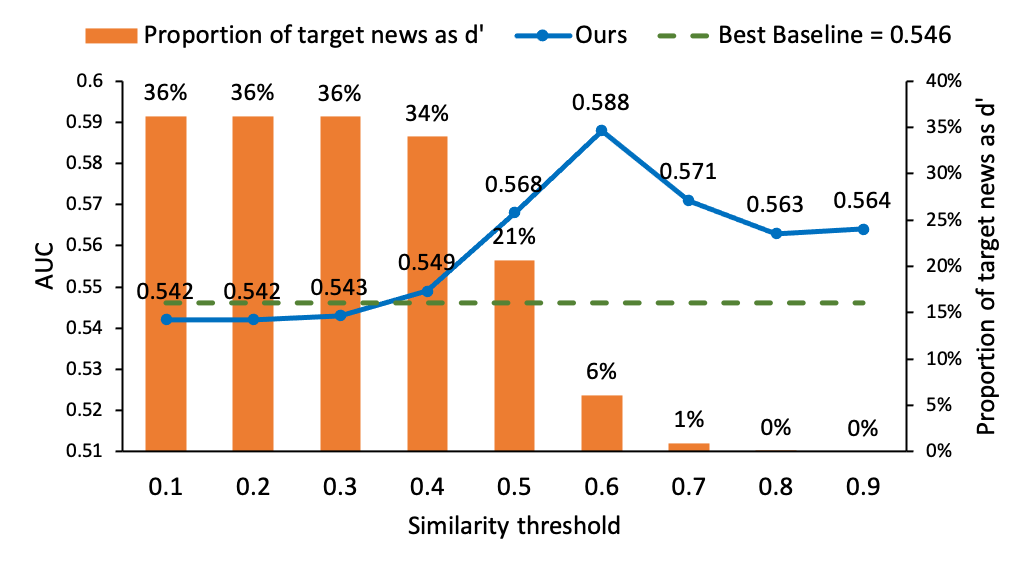}
\end{center}
\vspace{-0.4cm}
\caption{Influence of the similarity threshold  for selecting the most similar target domain news.}
\label{similarity_threshold}\vspace{-0.3cm}
\end{figure}

In the Cross-domain Extension module, we use a similarity threshold to control the selection of the most similar target domain news. The above reported results are from the setting of this threshold = 0.6, based on a grid search process. Here, we explore the influence of this similarity threshold when varying it from 0.1 to 0.9 in the MIND -> Adressa 4-shot setting. In Figure \ref{similarity_threshold}, the x-axis denotes the similarity threshold. The primary y-axis on the left denotes the AUC and the secondary y-axis on the right denotes the proportion of the target domain news selected as the most similar news $d^{\prime}$. We find that if the similarity threshold is small (0.1$\sim$0.4), although over 30\% target news are selected as $d^{\prime}$ to participate  in training, the recommendation model does not have a good performance.  This is because the selected target news $d^{\prime}$ are not actually similar to source news $d$. The low threshold makes the unrelated news pass through training.  With the increase of the similarity threshold, less but truly similar target domain news are used as $d^{\prime}$ for training. The threshold = 0.6 is the best setting, which introduces 6\% of the target domain news in training. 
Note that if the threshold is large (0.7$\sim$0.9),  there are only few  similar news found between the source domain and target domain. Although our method then has downgraded performance, it can still outperform the best baseline (Bilingual (Source+Trans.+Target)), demonstrating the effectiveness of our method on taking advantage of the introduced similar target news.

\section{Top-K Selection in Cross-domain Extension}
\begin{table}[]
\Description{Influence of Top-K Selection in Cross-domain Extension.}
\centering
\caption{Influence of Top-K Selection in Cross-domain Extension.}\vspace{-0.3cm}
\scalebox{0.95}{
\begin{tabular}{ll|l}
\hline
\multicolumn{2}{l|}{Top-K Selection}                                                                                                                     & AUC            \\ \hline
\multicolumn{2}{l|}{Top-1}                                                                                                                       & 0.588$\pm$0.020         \\ \hline

\multicolumn{2}{l|}{Top-2}                                                                                                                       & 0.579$\pm$0.015         \\ \hline
\multicolumn{2}{l|}{Top-3}                                                                                                                       & 0.574$\pm$0.013         \\ 
\hline
\end{tabular}
}
\label{top-k}\vspace{-0.3cm}
\end{table}

In the Cross-domain Extension module, we select the most (Top-1) similar target domain news as the augmentation. Here, we explore the influence of the Top-K selection when selecting Top-1, Top-2, Top-3 similar target domain news in the MIND->Adressa 4-shot setting. As shown in Table \ref{top-k}, the AUC of Top-2 and Top-3 selection are lower than Top-1 selection. This maybe because there is a large content gap between MIND news published in 2020 and Adressa news published in 2017. Selecting more target domain news can potentially introduce irrelevant news and have negative impact on the prediction performance.

\section{Additional Experimental Results}
\subsection{NDCG and MRR for zero-shot and few-shot transfer}
Here we report the NDCG@5, NDCG@10 and MRR performance of our methods and all baselines in both MIND -> Adressa and Adressa -> MIND zero-shot and few-shot setting. As shown in Table \ref{other_metric}, our method achieves better performance in all metrics and all settings.
\begin{table*}[t]
\Description{Zero-shot and few-shot recommendation performance comparison. 
The best result is in bold font, and the best  baseline is underlined. The unsolvable zero-shot case is indicated by $-$. Each experiment is repeated for 5 times and the mean and the standard deviation are reported.}
\centering
\caption{Zero-shot and few-shot recommendation performance comparison. 
The best result is in bold font, and the best  baseline is underlined. The unsolvable zero-shot case is indicated by $-$. Each experiment is repeated for 5 times and the mean and the standard deviation are reported.}
\label{tab2} \vspace{-0.4cm}
\begin{subtable}{1\textwidth}
\caption{Metric: NDCG@5}
\vspace{-0.2cm}
\resizebox{\textwidth}{!}{
\begin{tabular}{ll|lll|lll}
\hline
\multicolumn{2}{l|}{Cross-domain Setting}                                           & \multicolumn{3}{c|}{MIND -> Adressa}                                                                                        & \multicolumn{3}{c}{Adressa -> MIND}                                                                                          \\ \hline
\multicolumn{2}{l|}{training data available in  target domain}        & 0-shot                                  & 2-shot                                  & 4-shot                                 & 0-shot                                  & 2-shot                                  & 4-shot                                  \\ \hline
\multicolumn{2}{l|}{NRMS (Only Target)~\cite{wu-etal-2019-neural-news}}             & 0.552$\pm$0.009   & \underline{0.616$\pm$0.005}   & 0.601$\pm$0.004   & 0.247$\pm$0.008   & 0.274$\pm$0.004   & 0.263$\pm$0.011   \\
\multicolumn{2}{l|}{ZERO-SHOT (Source + Target)~\cite{zhang-etal-2021-cross}}       & 0.574$\pm$0.009   & 0.602$\pm$0.013   & 0.587$\pm$0.008   & 0.271$\pm$0.012   & 0.298$\pm$0.020   & 0.279$\pm$0.010   \\
\multicolumn{2}{l|}{Trans.-Align (Trans. + Target)~\cite{zhang-etal-2021-cross}}    & \underline{0.579$\pm$0.012}   & 0.615$\pm$0.006   & 0.600$\pm$0.010   & \underline{0.289$\pm$0.006}   & \underline{0.321$\pm$0.008}   & 0.298$\pm$0.004   \\
\multicolumn{2}{l|}{Bilingual (Source+Trans.+Target)~\cite{zhang-etal-2021-cross}}  & 0.577$\pm$0.010   & 0.607$\pm$0.014   & \underline{0.608$\pm$0.014}   & 0.288$\pm$0.009   & 0.314$\pm$0.008   & \underline{0.303$\pm$0.005}   \\
\multicolumn{2}{l|}{TDAR~\cite{Yu2020a}}            &$-$  & 0.581$\pm$0.005   & 0.576$\pm$0.010    &$-$   & 0.277$\pm$0.007   & 0.280$\pm$0.010    \\ \hline
\multicolumn{1}{c|}{\multirow{2}{*}{Ours}} & Random Masking            
& 0.588$\pm$0.005   & 0.625$\pm$0.009   & 0.625$\pm$0.005   & \textbf{0.308$\pm$0.007}   & 0.323$\pm$0.014   & \textbf{0.312$\pm$0.009}   \\
\multicolumn{1}{c|}{}                      & Random Masking+News-Align
& \textbf{0.598$\pm$0.010}   & \textbf{0.633$\pm$0.013}   & \textbf{0.635$\pm$0.015}   & 0.306$\pm$0.005   & \textbf{0.334$\pm$0.009}   & \textbf{0.312$\pm$0.006}  \\ \hline
\end{tabular}
}
\end{subtable}
\vspace{+0.1cm}
\begin{subtable}{1\textwidth}
\caption{Metric: NDCG@10}
\vspace{-0.2cm}
\resizebox{\textwidth}{!}{
\begin{tabular}{ll|lll|lll}
\hline
\multicolumn{2}{l|}{Cross-domain Setting}                                           & \multicolumn{3}{c|}{MIND -> Adressa}                                                                                        & \multicolumn{3}{c}{Adressa -> MIND}                                                                                          \\ \hline
\multicolumn{2}{l|}{training data available in  target domain}        & 0-shot                                  & 2-shot                                  & 4-shot                                 & 0-shot                                  & 2-shot                                  & 4-shot                                  \\ \hline
\multicolumn{2}{l|}{NRMS (Only Target)~\cite{wu-etal-2019-neural-news}}             & 0.642$\pm$0.008   & 0.685$\pm$0.005   & 0.682$\pm$0.004   & 0.310$\pm$0.006   & 0.344$\pm$0.005   & 0.324$\pm$0.008   \\
\multicolumn{2}{l|}{ZERO-SHOT (Source + Target)~\cite{zhang-etal-2021-cross}}       & 0.656$\pm$0.006   & 0.675$\pm$0.010   & 0.664$\pm$0.006   & 0.327$\pm$0.011   & 0.357$\pm$0.021   & 0.345$\pm$0.005   \\
\multicolumn{2}{l|}{Trans.-Align (Trans. + Target)~\cite{zhang-etal-2021-cross}}    & \underline{0.665$\pm$0.011}   & \underline{0.687$\pm$0.005}   & 0.681$\pm$0.007   & \underline{0.352$\pm$0.004}   & \underline{0.377$\pm$0.012}   & 0.359$\pm$0.004   \\
\multicolumn{2}{l|}{Bilingual (Source+Trans.+Target)~\cite{zhang-etal-2021-cross}}  & 0.663$\pm$0.008   & 0.681$\pm$0.013   & \underline{0.688$\pm$0.014}   & 0.346$\pm$0.006   & 0.375$\pm$0.006   & \underline{0.361$\pm$0.007}   \\
\multicolumn{2}{l|}{TDAR~\cite{Yu2020a}}           
  &$-$    & 0.660$\pm$0.005   & 0.662$\pm$0.008   &$-$    & 0.342$\pm$0.008   & 0.343$\pm$0.008   \\ \hline
\multicolumn{1}{c|}{\multirow{2}{*}{Ours}} & Random Masking            
& 0.672$\pm$0.004   & 0.695$\pm$0.007   & 0.702$\pm$0.003   & 0.364$\pm$0.004   & 0.383$\pm$0.011   & \textbf{0.371$\pm$0.006}   \\
\multicolumn{1}{c|}{}                      & Random Masking+News-Align
& \textbf{0.679$\pm$0.009}   & \textbf{0.698$\pm$0.010}   & \textbf{0.708$\pm$0.013}   & \textbf{0.365$\pm$0.004}   & \textbf{0.389$\pm$0.006}   & 0.370$\pm$0.005   \\ \hline
\end{tabular}
}
\end{subtable}
\vspace{+0.1cm}
\begin{subtable}{1\textwidth}
\caption{Metric: MRR}
\vspace{-0.2cm}
\resizebox{\textwidth}{!}{
\begin{tabular}{ll|lll|lll}
\hline
\multicolumn{2}{l|}{Cross-domain Setting}                                           & \multicolumn{3}{c|}{MIND -> Adressa}                                                                                        & \multicolumn{3}{c}{Adressa -> MIND}                                                                                          \\ \hline
\multicolumn{2}{l|}{training data available in  target domain}        & 0-shot                                  & 2-shot                                  & 4-shot                                 & 0-shot                                  & 2-shot                                  & 4-shot                                  \\ \hline
\multicolumn{2}{l|}{NRMS (Only Target)~\cite{wu-etal-2019-neural-news}}             & 0.381$\pm$0.009   & 0.436$\pm$0.005   & 0.426$\pm$0.006   & 0.244$\pm$0.006   & 0.271$\pm$0.002   & 0.257$\pm$0.007   \\
\multicolumn{2}{l|}{ZERO-SHOT (Source + Target)~\cite{zhang-etal-2021-cross}}       & 0.397$\pm$0.007   & 0.423$\pm$0.012   & 0.404$\pm$0.006   & 0.260$\pm$0.011   & 0.281$\pm$0.019   & 0.270$\pm$0.007   \\
\multicolumn{2}{l|}{Trans.-Align (Trans. + Target)~\cite{zhang-etal-2021-cross}}    & \underline{0.408$\pm$0.013}   & \underline{0.437$\pm$0.006}   & 0.425$\pm$0.007   & \underline{0.280$\pm$0.005}   & \underline{0.305$\pm$0.011}   & \underline{0.284$\pm$0.005}   \\
\multicolumn{2}{l|}{Bilingual (Source+Trans.+Target)~\cite{zhang-etal-2021-cross}}  & 0.404$\pm$0.008   & 0.429$\pm$0.015   & \underline{0.433$\pm$0.016}   & 0.273$\pm$0.007   & 0.303$\pm$0.007   & 0.281$\pm$0.006   \\
\multicolumn{2}{l|}{TDAR~\cite{Yu2020a}}           
  &$-$    & 0.407$\pm$0.007   & 0.403$\pm$0.008    &$-$    & 0.264$\pm$0.008   & 0.270$\pm$0.007   \\ \hline
\multicolumn{1}{c|}{\multirow{2}{*}{Ours}} & Random Masking            
& 0.415$\pm$0.005   & 0.445$\pm$0.008   & 0.448$\pm$0.004   & \textbf{0.295$\pm$0.004}   & 0.309$\pm$0.013   & 0.303$\pm$0.005   \\
\multicolumn{1}{c|}{}                      & Random Masking+News-Align
& \textbf{0.422$\pm$0.009}   & \textbf{0.449$\pm$0.012}   & \textbf{0.455$\pm$0.014}   & 0.294$\pm$0.003   & \textbf{0.317$\pm$0.005}   & \textbf{0.304$\pm$0.004}   \\ \hline
\end{tabular}
}
\end{subtable}
\label{other_metric}
\end{table*}

\subsection{NDCG and MRR for experiments with different amounts of users from the target domain}
\begin{figure*}[h]
\Description{NDCG@5, NDCG@10 and MRR when varying the amount of users involved in training from the target domain.}
\begin{center}
  \includegraphics[width=1\linewidth]{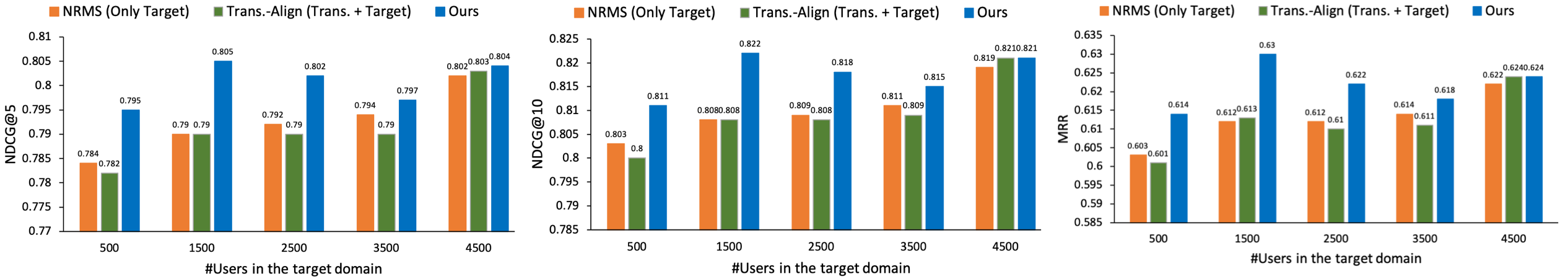}
\end{center}
 \vspace{-0.2cm}
\caption{NDCG@5, NDCG@10 and MRR when varying the amount of users involved in training from the target domain.}
\label{user_increase_ndcg5}\vspace{-0.1cm}
\end{figure*}

As shown in Figure \ref{user_increase_ndcg5}, the NDCG and MRR performance of our method are better than other baselines in the different amounts of users from the target domain setting. 

\subsection{NDCG and MRR for experiments with the Multilingual Pretrained Language Model}
\begin{figure*}[htbp]
\Description{Experiments with the Multilingual Pretrained Language Model in terms of NDCG@5, NDCG@10 and MRR.}
\begin{center}
  \includegraphics[width=1\linewidth]{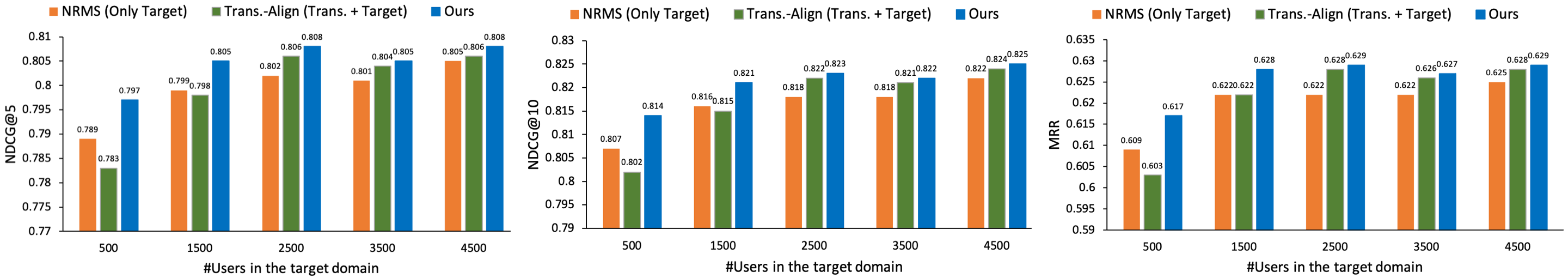}
\end{center}
 \vspace{-0.2cm}
\caption{Experiments with the Multilingual Pretrained Language Model in terms of NDCG@5, NDCG@10 and MRR.}
\label{plm_ndcg5}\vspace{-0.2cm}
\end{figure*}

As shown in Figure \ref{plm_ndcg5}, the NDCG and MRR performance of our method are better than other baselines in the experiments with the Multilingual Pretrained Language Model.

\end{document}